%% file: ms.tex
\documentclass[sigconf]{acmart}

\usepackage{graphicx}
\usepackage{subfiles}

\usepackage{todonotes}
\usepackage{booktabs}
\PassOptionsToPackage{hyphens}{url}\usepackage{hyperref}
\usepackage{multirow}

\usepackage{dsfont}
\usepackage{amsmath}

\usepackage{caption}
\usepackage{subcaption}

\usepackage{xcolor}

\AtBeginDocument{%
  \providecommand\BibTeX{{%
    \normalfont B\kern-0.5em{\scshape i\kern-0.25em b}\kern-0.8em\TeX}}}

\copyrightyear{2024}
\acmYear{2024}
\setcopyright{rightsretained}
\acmConference[SIGIR '24]{Proceedings of the 47th International ACM SIGIR Conference on Research and Development in Information Retrieval}{July 14--18, 2024}{Washington, DC, USA}
\acmBooktitle{Proceedings of the 47th International ACM SIGIR Conference on Research and Development in Information Retrieval (SIGIR '24), July 14--18, 2024, Washington, DC, USA}
\acmDOI{10.1145/3626772.3657955}
\acmISBN{979-8-4007-0431-4/24/07}

\begin{document}

\title{Distillation for Multilingual Information Retrieval}

\author{Eugene Yang}
\affiliation{%
  \institution{HLTCOE, Johns Hopkins University}
  \city{Baltimore}
  \state{Maryland}
  \country{USA}
}
\email{eugene.yang@jhu.edu}

\author{Dawn Lawrie}
\affiliation{%
  \institution{HLTCOE, Johns Hopkins University}
  \city{Baltimore}
  \state{Maryland}
  \country{USA}
}
\email{lawrie@jhu.edu}

\author{James Mayfield}
\affiliation{%
  \institution{HLTCOE, Johns Hopkins University}
  \city{Baltimore}
  \state{Maryland}
  \country{USA}
}
\email{mayfield@jhu.edu}

\begin{abstract}
Recent work in cross-language information retrieval (CLIR),
where queries and documents are in different languages,
has shown the benefit of the Translate-Distill framework that trains a cross-language neural dual-encoder model
using translation and distillation. 
However, Translate-Distill only supports a single document language. 
Multilingual information retrieval (MLIR),
which ranks a multilingual document collection,
is harder to train than CLIR
because the model must assign comparable relevance scores to documents in different languages.
This work extends Translate-Distill and propose Multilingual Translate-Distill (MTD) for MLIR. 
We show that ColBERT-X models trained with MTD outperform their counterparts trained with Multilingual Translate-Train,
which is the previous state-of-the-art training approach,
by 5\% to 25\% in nDCG@20 and 15\% to 45\% in MAP. 
We also show that the model is robust to the way languages are mixed in training batches. 
Our implementation is available on GitHub.
\end{abstract}



\maketitle

\input{_mixing_figures}
\input{1_intro}

\input{2_background}
\input{3_method}

\input{4_exp}
\input{5_results}
\input{6_conclusion}

\bibliographystyle{ACM-Reference-Format}
\bibliography{sample-base}

\end{document}

%% file: _mixing_figures.tex
\begin{figure*}[t]
     \centering
     \begin{subfigure}[b]{0.3\textwidth}
         \centering
         \includegraphics[width=0.88\textwidth]{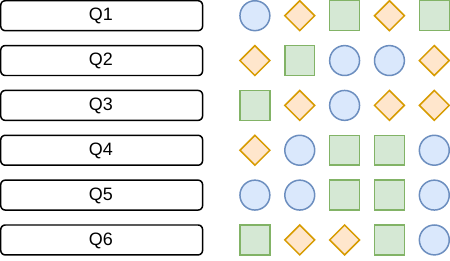}
         \caption{Mix Passages}\label{fig:strategies-mixpass}
     \end{subfigure}
     \hfill
     \begin{subfigure}[b]{0.3\textwidth}
         \centering
         \includegraphics[width=0.88\textwidth]{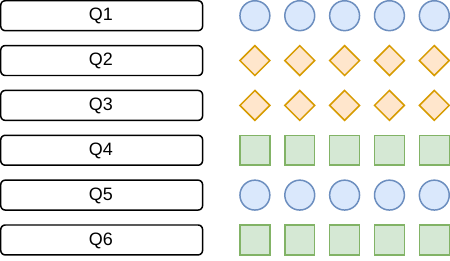}
         \caption{Mix Entries}\label{fig:strategies-mixentries}
     \end{subfigure}
     \hfill
     \begin{subfigure}[b]{0.3\textwidth}
         \centering
         \includegraphics[width=0.88\textwidth]{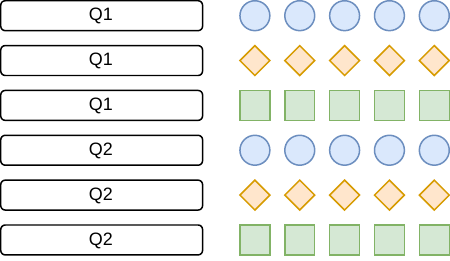}
         \caption{Round Robin Entries}\label{fig:strategies-roundrobin}
     \end{subfigure}
\vspace{-1em}
\caption{Three language mixing strategies for Multilingual Translate-Distill. Each row indicates an entry with a query and a list of sampled passages in the training mini-batch. Circles, diamonds, and squares represent different document languages. }\label{fig:strategies}
\end{figure*}

%% file: 1_intro.tex
\section{Introduction}

We define Multilingual Information Retrieval (MLIR)
as search over a \textit{multilingual} collection of \textit{monolingual} documents to produce a single ranked list~\cite{peters2012multilingual, si2008effective, tsai2008study, rahimi2015multilingual, mtt}. 
The retrieval system must retrieve and rank documents based only on query relevance,
independent of document language.
This is challenging in part because cross-language systems may be unable to exploit surface forms.
Our evaluation uses CLEF data~\cite{clef2003} with English queries and French, German, Spanish, and English documents; 
CLEF data ~\cite{clef2000, clef2001, clef2002, clef2003} with English queries and French, German, and Italian documents;  
and TREC NeuCLIR data~\cite{neuclir2022, neuclir2023} with English queries and Chinese, Persian, and Russian documents. 

Dual-encoder retrieval models
such as ColBERT~\cite{colbert} that matches token embeddings,
and DPR~\cite{dpr} that matches query and document embeddings,
have shown good results in both monolingual~\cite{colbertv2}
and cross-language~\cite{colbertx, zhang2021mrtydi, tdistill, li2022learning} retrieval.
These approaches use pre-trained language models like multilingual BERT~\cite{bert} and XLM-RoBERTa~\cite{conneau2020xlmr}
as text encoders to place queries and documents into a joint semantic space;
this allows embedding distances to be calculated across languages. 
Multilingual encoders are generally trained monolingually on multiple languages~\cite{bert, xlmr},
which leads to limited cross-language ability.
Therefore, careful fine-tuning,
such as Translate-Train~\cite{colbertx}, C3 Pretraining~\cite{c3} and Native-Train~\cite{blade},
are essential to be able to match across languages~\cite{shi2019cross, tdistill, li2022learning}. 

Generalizing from one to multiple document languages is not trivial. 
Prior work showed that Multilingual Translate-Train (MTT) \cite{mtt} of ColBERT-X
using
training data translated into all document languages
is more effective than BM25 search over documents translated into the query language.
Searching translated documents with the English ColBERT model is even more effective than MTT,
but incurs a high translation cost
at indexing time
compared to MTT's amortized cost of translating the training corpus. 
This work aims to
develop MLIR training that produces more effective models than its monolingual English counterparts. 

Knowledge distillation has shown success monolingually~\cite{rocketqa, colbertv2, formal2021splade},
so we adapt this concept to train MLIR models. 
In Translate-Distill~\cite{tdistill}, a way to train CLIR ColBERT-X models,
a teacher model scores monolingual training data using text in whichever language produces its best results. 
Then when training the student ColBERT-X model,
training data is translated into the languages that match the final CLIR task.
That work showed that the student model is on par with or more effective than a retrieve-and-rerank system
that uses that same teacher model as a reranker. 
We propose Multilingual Translate-Distill (MTD), a multilingual generalization of Translate-Distill. 
Instead of training with a single document language,
we translate training passages into all document languages. 
This opens a design space of how to mix languages in training batches. 

This paper contributes
(1) an effective training approach for an MLIR dual-encoder that combines translation and distillation; 
(2) models trained with MTD that are more effective than the previously reported state-of-the-art MLIR model,
ColBERT-X trained with MTT; and 
(3) a robustness analysis of mini-batch passage mixing strategies. 
Models and implementation are available on Huggingface Models\footnote{\url{https://huggingface.co/collections/hltcoe/multilingual-translate-distill-66280df75c34dbbc1708a22f}} and GitHub.\footnote{\url{https://github.com/hltcoe/colbert-x}}

%% file: 2_background.tex
\section{Background}

An IR problem can be ``multilingual'' in several ways. 
For example, \citet{hull1996querying} described a multilingual IR problem of monolingual retrieval in multiple languages, as in~\citet{blloshmi2021ir},
or alternatively, multiple
CLIR tasks %
in several languages~\cite{hc4,clef2001,clef2002,clef2003,ntcir2007overview}.
We adopt the Cross-Language Evaluation Forum (CLEF)'s notion of MLIR:
using a query to construct one ranked list across documents in several languages~\cite{peters2002importance}. 
We acknowledge that this definition excludes mixed-language or code-switched queries and documents,
other cases to which ``multilingual'' has been applied.  

Prior to neural retrieval, MLIR systems generally relied on cross-language dictionaries
or machine translation  models~\cite{darwish2003probabilistic, kraaij2003embedding, mcnamee2002comparing}.
Translating documents into the query language casts MLIR as monolingual in that language~\cite{magdy2011should, granell2014multilingual, rahimi2015multilingual}.
While translating queries into each document language is almost always computationally more economical than translating the documents,
it casts the MLIR problem as multiple monolingual problems
whose results must be merged to form the final MLIR ranked list~\cite{peters2012multilingual, si2008effective, tsai2008study}. 
Moreover, quality differences between translation models could bias results by systematically
ranking documents in some languages higher~\cite{mtt, huang2023soft}.

Recent work in representation learning for IR~\cite{formal2021splade, cocondenser, reimers2019sentence}
and fast dense vector search algorithms~\cite{hnsw, jegou2010product, faiss}
spawned a new class of models called dual-encoders. 
These models encode queries and documents simultaneously into one or more dense vectors
representing tokens, spans, or entire sequences~\cite{colbert, li2023citadel, li2023slim, dpr}. 
While replacing the underlying language model with a multilingual one,
such as multilingual BERT~\cite{bert} and XLM-RoBERTa~\cite{xlmr},
produces systems that accept queries and documents in multiple languages,
zero-shot transfer of a model trained only monolingually to a CLIR or MLIR problem is suboptimal;
it leads to systems even less effective than BM25 over document translations~\cite{colbertx, mtt}. 
Therefore, designing an effective fine-tuning process for transforming multilingual language models into multilingual IR models is critical. 

Various retrieval fine-tuning approaches have been explored,
such as contrastive learning~\cite{dpr, colbert, colbertv2},
hard-negative mining~\cite{hofstatter2021efficiently, formal2021splade},
and knowledge distillation~\cite{formal2021splade, colbertv2, rocketqa}. 
Knowledge distillation has demonstrated more effective results in both monolingual and cross-language IR~\cite{li2022learning, tdistill} than the others.
The recently proposed Translate-Distill approach decoupled the input languages of the teacher and student models.
This allowed large English rerankers to train ColBERT-X for CLIR,
leading to state-of-the-art CLIR effectiveness measured on the NeuCLIR 22 benchmark~\cite{neuclir2022}. 
Recent work by \citet{huang2023soft} proposes a language-aware decomposition for prompting (or augmenting) the document encoder. 
In this work, we explore the simple idea of relying on translations of MS MARCO and distilling the ranking knowledge from a large MonoT5 model with mT5XXL underneath~\cite{unicamp-at-neuclir, monot5, mt5}. 

%% file: 3_method.tex
\section{Multilingual Translate-Distill}

Our proposed Multilingual Transalte-Distill (MTD) training approach
requires a monolingual training corpus consisting of queries and passages;
no relevance labels are required. 

\subsection{Knowledge Distillation}

To train a student dual-encoder model for MLIR,
we first use two teacher models:
a query-passage selector and a query-passage scorer.
Following \citet{tdistill}, the query-passage selector retrieves $k$ passages for each query. 
This can be replaced by any hard-negative mining approach~\cite{rocketqa, hofstatter2021efficiently} or by adapting publicly available mined passages.\footnote{For example, \url{https://huggingface.co/datasets/sentence-transformers/msmarco-hard-negatives}.}
The query-passage scorer then scores each query-passage pair with high accuracy. 
The scorer is essentially a reranker from which we would like to distill ranking knowledge implicit in an expensive model such as MonoT5~\cite{monot5}
that is generally too slow to apply by itself. 
The final product from the two teachers is a set of tuples,
each containing a query, a passage, and the associated teacher score. 
We use these data
to train the student dual-encoder model.
Specifically, for each training mini-batch of size $n$,
we select $n$ training queries and sample $m$ retrieved passage IDs.
To teach the student model to rank documents across languages,
we translate each passage into all of the target languages.
When constructing the mini-batch, we determine the language for each passage ID, which we discuss in more detail in the next section.
Finally, the loss function is the KL divergence between the teacher and student scores
on the query and the translated passages. 

\subsection{Language Mixing Strategies}

To train an effective ColBERT-X model for MLIR,
each training batch must include documents in more than one language~\cite{mtt}. 
Training with MTD opens a design space for selecting languages for the mini-batch passages. 
We experiment with three mixing strategies (see Figure~\ref{fig:strategies}):

\textbf{Mix Passages.} In each training batch entry,
all passages are randomly assigned to one of the document languages.
In this case, each language is equally likely to be present during training. 
Each language also has an equal probability of being assigned to any passage in such a way that language representation is balanced, thus a language is just as likely to be assigned to a passage with a high score as a low score.  
This mixing method directly trains the student model to rank passages in different languages. 

\textbf{Mix Entries.} Alternatively, we can assign the same randomly selected language to all passages associated with a query.
This method ensures the translation quality does not become a possible feature that the student model could rely on if there is a language with which the machine translation model struggles.
While not directly learning MLIR,
this model jointly learns multiple CLIR tasks
with distillation and eventually learns the MLIR task.

\textbf{Round Robin Entries.} To ensure the model equally learns the ranking problem for all languages,
we experiment with training query repetition to present passages from all languages. 
In this case, the model learns the CLIR tasks using the same set of queries instead of a random subset
when mixing entries. 
However, this reduces the number of queries per mini-batch given some fixed GPU memory size. 
Given this memory constraint, round robin may not be feasible if the number of document languages exceeds the number of entries the GPU can hold at once.

%% file: 4_exp.tex
\section{Experiments}

\begin{table}[t]
\caption{Collection Statistics}\label{tab:collections}
\vspace{-1em}
\centering
\begin{tabular}{l|cc|cc}
\toprule
                &     \multicolumn{2}{c|}{CLEF}  &  \multicolumn{2}{c}{NeuCLIR} \\
                &  Subset\cite{huang2023soft}  & 2003 &     2022 &    2023 \\
\midrule
Languages       &   de, fr, it &  de, fr, es, en &  \multicolumn{2}{c}{zh, fa, ru} \\
\midrule
\# of Docs      &        0.24M &           1.05M &  \multicolumn{2}{c}{10.04M} \\
\# of Passages  &        1.90M &           6.96M &  \multicolumn{2}{c}{58.88M} \\
\midrule
\# of Topics    &          113 &              60 &       41 &      65 \\
Avg. Rel/Topic  &        40.73 &          102.42 &   125.46 &   67.77 \\
\bottomrule
\end{tabular}
\end{table}

We evaluate our proposed model on four MLIR evaluation collections: 
a subset of CLEF00-03 curated by \citet{huang2023soft}\footnote{The collection is reconstructed by using the author-provided document IDs, which excludes a large portion of unjudged documents.
Documents added in subsequent years are also excluded. Thus some judged relevant documents are also excluded.};
CLEF03 with German, French, Spanish, and English~\cite{clef2003};
and NeuCLIR 2022~\cite{neuclir2022} and 2023~\cite{neuclir2023}. 
Collection statistics are summarized in Table~\ref{tab:collections}. 
Queries %
are English titles concatenated with descriptions. 

We use MS MARCO~\cite{msmarco} to train the MLIR ColBERT-X models with MTD,
for which we adopt the PLAID-X implementation released by \citet{tdistill}.\footnote{\url{https://github.com/hltcoe/ColBERT-X}}
We use the English ColBERTv2 model released by \citet{colbertv2} that was also trained with knowledge distillation\footnote{\url{https://huggingface.co/colbert-ir/colbertv2.0}} 
and MonoT5 with mT5XXL released by \citet{unicamp-at-neuclir}\footnote{\url{https://huggingface.co/unicamp-dl/mt5-13b-mmarco-100k}} as query-passage selector and scorer, respectively. 
Both the selector and the scorer received English MS MARCO queries and passages to generate training teacher scores. 

To support MTD training, we translated the MS MARCO passages with Sockeye v2~\cite{sockeye2amta, sockeye2whitepaper} into the document languages. 
Student ColBERT-X models are fine-tuned from the XLM-RoBERTa large models~\cite{xlmr} using 8 NVidia V100 GPUs (32GB memory) 
for 200,000 gradient steps with a mini-batch size of 8 entries each associated with 6 passages on each GPU. 
We use AdamW optimizer with a $5\times 10^{-6}$ learning rate and half-precision floating points. 

\input{_main_result_table}

Documents are split into 180 token passages with a stride of 90 before indexing. 
The number of resulting passages is reported in Table~\ref{tab:collections}.
We index the collection with PLAID-X using one residual bit. 
At search time, PLAID-X retrieves passages, and document scores are aggregated using MaxP~\cite{maxp}. 
For each query, we return the top 1000 documents for evaluation. 

To demonstrate MTD effectiveness,
we report baseline ColBERT models that are trained differently:
English ColBERT~\cite{colbertv2},
ColBERT-X with Multilingual Translate-Train (MTT)~\cite{mtt},
and ColBERT-X with English Distillation (ED). 
Since English ColBERT does not accept text in other languages,
we index the collection with documents machine-translated into English
(marked ``DT'' in Table~\ref{tab:main-results}). 
ColBERT-X models trained with MTT use the training triples released by MS MARCO
with hyperparameters similar to the MTD ones except for the number of queries per batch per GPU is increased to 32. 
Finally, the English Distillation models are only exposed to English queries and passages during fine-tuning instead of the translated text. 
It performs a zero-shot language transfer at indexing and search time.

We also compare our models to the recently published KD-SPD~\cite{huang2023soft},
which is a language-aware MLIR model that encodes the entire text sequence as a single vector.
To provide a broader context,
we report sparse retrieval baselines PSQ-HMM~\cite{darwish2003probabilistic, xu2000cross, yang2024psq} and BM25 with translated documents,
which are two strong MLIR baselines reported in NeuCLIR 2023~\cite{neuclir2023}. 

We report nDCG@20, MAP, and Recall at 1000 for the CLEF03 and NeuCLIR collections. 
To enable comparison to \citet{huang2023soft}, we report nDCG@10, MAP@100, and Recall@100 on the CLEF00-03 subset. 
To test statistical superiority between two systems,
we use a one-sided paired t-test with 95\% confidence on the per-topic metric values. 
When testing for statistical ``equivalence'' where the null hypothesis is that the effectiveness of the two systems differ,
we use a paired Two One-sided T-Tests (TOST)~\cite{lakens2017equivalence, schuirmann1987comparison} with a threshold of 0.05 and 95\% confidence. 

%% file: _main_result_table.tex
\begin{table*}[t]
\setlength\tabcolsep{0.27em}
\renewcommand{\arraystretch}{1.15}

\caption{MLIR system effectiveness. 
Numbers in superscripts indicate the system of the row is statistically better than the systems in the superscript with 95\% confidence by conducting a one-sided paired t-test.  
Numbers in subscripts indicate the system of the row is statistically \textit{identical} within 0.05 in value to the systems in the subscripts with 95\% confidence by conducting paired TOSTs. 
Bonferroni corrections are applied to both sets of statistical tests. 
}\label{tab:main-results}
\vspace{-1em}

\resizebox{\textwidth}{!}{
\centering
\begin{tabular}{p{1em}p{9em}|lll|lll|lll|lll}
\toprule
{}  & {} & \multicolumn{3}{c|}{CLEF00-03 Subset~\cite{huang2023soft}}  & \multicolumn{3}{c|}{CLEF 2003} & \multicolumn{3}{c|}{NeuCLIR 2022 MLIR} & \multicolumn{3}{c}{NeuCLIR 2023 MLIR} \\
{}  &             Measure &  \multicolumn{1}{c}{nDCG} &  \multicolumn{1}{c}{MAP} &  \multicolumn{1}{c|}{Recall} &
                             \multicolumn{1}{c}{nDCG} &  \multicolumn{1}{c}{MAP} &  \multicolumn{1}{c|}{Recall} &
                             \multicolumn{1}{c}{nDCG} &  \multicolumn{1}{c}{MAP} &  \multicolumn{1}{c|}{Recall} &
                             \multicolumn{1}{c}{nDCG} &  \multicolumn{1}{c}{MAP} &  \multicolumn{1}{c}{Recall} \\
{}  &         Rank Cutoff &  \multicolumn{1}{c}{10} &  \multicolumn{1}{c}{100} &  \multicolumn{1}{c|}{100} &
                             \multicolumn{1}{c}{20} & \multicolumn{1}{c}{1000} & \multicolumn{1}{c|}{1000} &  
                             \multicolumn{1}{c}{20} & \multicolumn{1}{c}{1000} & \multicolumn{1}{c|}{1000} &  
                             \multicolumn{1}{c}{20} & \multicolumn{1}{c}{1000} & \multicolumn{1}{c}{1000} \\
\midrule
{}  & \multicolumn{13}{l}{Baselines} \\
\midrule
(0) & KD-SPD\cite{huang2023soft}  &          0.416 &                  0.220 &                  0.469 &                --  &                   --  &                   --  &              --  &                    --  &                    --  &                    -- &                    --  &                    -- \\
(1) & PSQ-HMM             &            0.529$^{0}$ &            0.339$^{0}$ &            0.617$^{0}$ &              0.445 &                 0.282 &                 0.711 &            0.315 &                  0.193 &                  0.594 &                 0.289 &                  0.225 &                  0.693 \\
(2) & DT >> BM25          &            0.568$^{0}$ &           0.388$^{01}$ &           0.662$^{01}$ &        0.636$^{1}$ &           0.453$^{1}$ &           0.857$^{1}$ &            0.338 &                  0.215 &                  0.633 &                 0.316 &                  0.275 &                  0.756 \\
(3) & DT >> ColBERT       &       0.609$^{01}_{4}$ &       0.422$^{01}_{4}$ &       0.700$^{01}_{4}$ &        0.669$^{1}$ &           0.497$^{1}$ &          0.889$^{14}$ &      0.403$^{1}$ &           0.285$^{12}$ &          0.708$^{124}$ &           0.361$^{1}$ &        0.298$^{1}_{4}$ &            0.786$^{1}$ \\
(4) & ColBERT-X MTT       &       0.613$^{01}_{3}$ &       0.411$^{01}_{3}$ &       0.687$^{01}_{3}$ &        0.643$^{1}$ &           0.451$^{1}$ &           0.827$^{1}$ &            0.375 &                  0.236 &                  0.612 &                 0.330 &        0.281$^{1}_{3}$ &                  0.760 \\
(5) & ColBERT-X ED        &      0.638$^{012}_{8}$ &  0.457$^{01234}_{678}$ &  0.732$^{01234}_{678}$ &  \textbf{0.699}$^{14}_{8}$ &  0.530$^{124}_{678}$ &   0.920$^{124}_{78}$ &      0.393 &                  0.263 &           0.687$^{14}$ &           0.357$^{1}$ &            0.317$^{1}$ &          0.827$^{124}$ \\
\midrule
{}  & \multicolumn{13}{l}{ColBERT-X MTD with Different Mixing Strategies} \\
\midrule
(6) & Mix Passages        &   0.666$^{01234}_{78}$ &  0.471$^{01234}_{578}$ &  0.747$^{01234}_{578}$ &        0.675$^{1}$ &    0.520$^{14}_{57}$ &     0.901$^{14}_{7}$ &     0.444$^{12}$ &   0.340$^{1245}_{78}$ &   0.762$^{1245}_{78}$ &  \textbf{0.404}$^{1245}_{78}$ &  0.367$^{12345}_{78}$ &  0.868$^{12345}_{78}$ \\
(7) & Mix Entries         &  \textbf{0.674}$^{012345}_{68}$ &  0.469$^{01234}_{568}$ &  0.745$^{01234}_{568}$ &        0.686$^{1}$ &    0.522$^{14}_{56}$ &  0.911$^{124}_{568}$ &   0.461$^{1245}$ &  \textbf{0.347}$^{12345}_{68}$ &  \textbf{0.768}$^{12345}_{68}$ &   0.397$^{124}_{68}$ &  \textbf{0.372}$^{12345}_{68}$ &  \textbf{0.877}$^{12345}_{678}$ \\
(8) & Round Robin Entries &  0.656$^{01234}_{567}$ &  \textbf{0.476}$^{01234}_{567}$ & \textbf{0.751}$^{012345}_{567}$ &  \textbf{0.699}$^{12}_{5}$ &   \textbf{0.535}$^{1234}_{5}$ &  \textbf{0.922}$^{1234}_{57}$ &  \textbf{0.474}$^{12345}$ &  0.341$^{12345}_{67}$ &   0.761$^{1245}_{67}$ &   0.388$^{124}_{67}$ &  0.347$^{12345}_{67}$ &   0.856$^{1234}_{67}$ \\

\bottomrule
\end{tabular}
}

\end{table*}

%% file: 5_results.tex
\section{Results}

\input{_joint_training_results}

Table~\ref{tab:main-results} summarizes our experiments. 
ColBERT-X models trained with MTD are more effective than those with MTT across all four evaluation collections,
demonstrating a 5\% (CLEF03 0.643 to 0.675 with mix passages)
to 26\% (NeuCLIR22 0.375 to 0.474 with round robin entries)
improvement in nDCG@20 and 15\% (CLEF03 0.451 to 0.520 with mix passages) to 47\% (NeuCLIR22 0.236 to 0.347 with mix entries) in MAP. 
MTD-trained ColBERT-X models over documents in their native form are significantly more effective than
translating all documents into English and searching with English ColBERT. 

Since the languages in the two CLEF collections are closer to English than those in NeuCLIR,
the ColBERT-X model trained with English texts (Row 5) 
still provides reasonable effectiveness using (partial) zero-shot language transfer during inference. 
MTD yields identical effectiveness to ED based on the TOST equivalence test in the two CLEF collections by measuring MAP (Table~\ref{tab:main-results}).
In contrast, NeuCLIR languages do not benefit from this phenomenon. 
Instead, training directly with text in document languages
enhances both the general language modeling and retrieval ability of the student models. 
In NeuCLIR 2022 and 2023, student ColBERT-X models trained with MTD (Rows 6 to 8) are 9\% (NeuCLIR23 0.317 to 0.347 with round robin entries) to 32\% (NeuCLIR22 0.263 to 0.347 with mix entries) more effective than ED (Row 5) by measuring MAP. 

\subsection{Ablation on Language Mixing Strategies}
Since the TOST equivalence tests show that the three mixing strategies demonstrate statistically similar MAP and Recall for all collections except for a few cases in CLEF 2003 
(CLEF 2003 may be an outlier because it has English documents, a known source of bias in MLIR~\cite{mtt}). 
We conclude that MTD is robust to how languages are mixed during training
as long as multiple languages are present in each training mini-batch~\cite{mtt}.
Such robustness provides operational flexibility to practitioners creating MLIR models. 
Since passage translation might not be available for all languages,
mixing passages allows selecting passages only from a subset of languages. 
Mixing entries also allows training entries to be filtered for specific languages
if relevance is known to drop after translation. 

When evaluating with nDCG@20, the differences are larger but less consistent. 
For the two CLEF collections and NeuCLIR 2022,
topics were developed for a single language before obtaining relevance judgments across all languages.
These topics may not be well-attested in all document languages,
resulting in some CLIR topics with few relevant documents. 
For these three collections, models trained with mixed CLIR tasks
(mix and round-robin entries) are more effective at the top of the ranking. 
High variation among topics leads to inconclusive statistical significance results,
suggesting opportunities for result fusion.
NeuCLIR 2023 topics were developed bilingually,
so topics are not socially or culturally tied to a single language;
this leads to statistically equivalent nDCG@20 results. 

\subsection{Training Language Ablation}

Finally, we explore training with languages beyond the ones in the document collection. 
Table~\ref{tab:joint-training} shows MTD-trained models for CLEF 2003, NeuCLIR, and both on each collection. 
Due to GPU memory constraints, we exclude the round-robin strategy from this ablation. 

We observe that models trained with the mix passages strategy
are more robust than the mix-entries variants when training on CLEF and evaluating on NeuCLIR and vice versa. This shows smaller degradation when facing language mismatch between training and inference. 
Surprisingly, training on NeuCLIR languages with the mix passage strategy yields numerically higher nDCG@20 than training on CLEF (0.675 to 0.688). 

When training both CLEF and NeuCLIR languages,
effectiveness is generally worse than only training on the evaluation languages. 
This trend suggests the models might be facing capability limits in the neural model,
or picking up artifacts from the quality differences in the translation. 
This observation demands more experimentation on MLIR dual-encoder models, which we leave for future work. 

%% file: _joint_training_results.tex
\begin{table}[t]
\caption{nDCG@20 on training with more languages}\label{tab:joint-training}
\vspace{-1em}
    \centering
    \begin{tabular}{ll|ccc}
\toprule
&                 & \multicolumn{3}{c}{Training Languages} \\
& Evaluation Collection& CLEF03 & NeuCLIR &    Both \\
\midrule
\multirow{3}{1cm}{Mix Passages}
& CLEF 2003         &   0.675 &   0.688 &   0.694 \\
& NeuCLIR 2022 MLIR &   0.437 &   0.444 &   0.431 \\
& NeuCLIR 2023 MLIR &   0.377 &   0.404 &   0.406 \\
\midrule
\multirow{3}{1cm}{Mix Entries}
& CLEF 2003         &   0.686 &   0.679 &   0.680 \\
& NeuCLIR 2022 MLIR &   0.424 &   0.461 &   0.445 \\
& NeuCLIR 2023 MLIR &   0.359 &   0.397 &   0.379 \\
\bottomrule

\end{tabular}

\end{table}

%% file: 6_conclusion.tex
\section{Conclusion}

We propose Multilingual Translate-Distill (MTD) for training MLIR dual-encoder models. 
We demonstrated that ColBERT-X models trained with the proposed MTD are more effective than using previously proposed MLIR training techniques on four MLIR collections. 
By conducting statistical equivalence tests, we showed that MTD is robust to the mixing strategies of the languages in the training mini-batch.

%% file: ms.bbl

\begin{thebibliography}{56}


\ifx \showCODEN    \undefined \def \showCODEN     #1{\unskip}     \fi
\ifx \showDOI      \undefined \def \showDOI       #1{#1}\fi
\ifx \showISBNx    \undefined \def \showISBNx     #1{\unskip}     \fi
\ifx \showISBNxiii \undefined \def \showISBNxiii  #1{\unskip}     \fi
\ifx \showISSN     \undefined \def \showISSN      #1{\unskip}     \fi
\ifx \showLCCN     \undefined \def \showLCCN      #1{\unskip}     \fi
\ifx \shownote     \undefined \def \shownote      #1{#1}          \fi
\ifx \showarticletitle \undefined \def \showarticletitle #1{#1}   \fi
\ifx \showURL      \undefined \def \showURL       {\relax}        \fi
\providecommand\bibfield[2]{#2}
\providecommand\bibinfo[2]{#2}
\providecommand\natexlab[1]{#1}
\providecommand\showeprint[2][]{arXiv:#2}

\bibitem[\protect\citeauthoryear{Blloshmi, Pasini, Campolungo, Banerjee,
  Navigli, and Pasi}{Blloshmi et~al\mbox{.}}{2021}]%
        {blloshmi2021ir}
\bibfield{author}{\bibinfo{person}{Rexhina Blloshmi}, \bibinfo{person}{Tommaso
  Pasini}, \bibinfo{person}{Niccol{\`o} Campolungo}, \bibinfo{person}{Somnath
  Banerjee}, \bibinfo{person}{Roberto Navigli}, {and}
  \bibinfo{person}{Gabriella Pasi}.} \bibinfo{year}{2021}\natexlab{}.
\newblock \showarticletitle{IR like a SIR: sense-enhanced information retrieval
  for multiple languages}. In \bibinfo{booktitle}{\emph{Proceedings of the 2021
  Conference on Empirical Methods in Natural Language Processing}}.
  \bibinfo{pages}{1030--1041}.
\newblock


\bibitem[\protect\citeauthoryear{Braschler}{Braschler}{2001a}]%
        {clef2000}
\bibfield{author}{\bibinfo{person}{Martin Braschler}.}
  \bibinfo{year}{2001}\natexlab{a}.
\newblock \showarticletitle{CLEF 2000 --- Overview of Results}. In
  \bibinfo{booktitle}{\emph{Cross-Language Information Retrieval and
  Evaluation}}, \bibfield{editor}{\bibinfo{person}{Carol Peters}} (Ed.).
  \bibinfo{publisher}{Springer Berlin Heidelberg}, \bibinfo{address}{Berlin,
  Heidelberg}, \bibinfo{pages}{89--101}.
\newblock
\showISBNx{978-3-540-44645-3}


\bibitem[\protect\citeauthoryear{Braschler}{Braschler}{2001b}]%
        {clef2001}
\bibfield{author}{\bibinfo{person}{Martin Braschler}.}
  \bibinfo{year}{2001}\natexlab{b}.
\newblock \showarticletitle{{CLEF} 2001—Overview of Results}. In
  \bibinfo{booktitle}{\emph{Workshop of the Cross-Language Evaluation Forum for
  European Languages}}. Springer, \bibinfo{pages}{9--26}.
\newblock


\bibitem[\protect\citeauthoryear{Braschler}{Braschler}{2002}]%
        {clef2002}
\bibfield{author}{\bibinfo{person}{Martin Braschler}.}
  \bibinfo{year}{2002}\natexlab{}.
\newblock \showarticletitle{{CLEF} 2002—Overview of results}. In
  \bibinfo{booktitle}{\emph{Workshop of the Cross-Language Evaluation Forum for
  European Languages}}. Springer, \bibinfo{pages}{9--27}.
\newblock


\bibitem[\protect\citeauthoryear{Braschler}{Braschler}{2003}]%
        {clef2003}
\bibfield{author}{\bibinfo{person}{Martin Braschler}.}
  \bibinfo{year}{2003}\natexlab{}.
\newblock \showarticletitle{{CLEF} 2003--Overview of results}. In
  \bibinfo{booktitle}{\emph{Workshop of the Cross-Language Evaluation Forum for
  European Languages}}. Springer, \bibinfo{pages}{44--63}.
\newblock


\bibitem[\protect\citeauthoryear{Conneau, Khandelwal, Goyal, Chaudhary, Wenzek,
  Guzm{\'a}n, Grave, Ott, Zettlemoyer, and Stoyanov}{Conneau
  et~al\mbox{.}}{2020a}]%
        {conneau2020xlmr}
\bibfield{author}{\bibinfo{person}{Alexis Conneau}, \bibinfo{person}{Kartikay
  Khandelwal}, \bibinfo{person}{Naman Goyal}, \bibinfo{person}{Vishrav
  Chaudhary}, \bibinfo{person}{Guillaume Wenzek}, \bibinfo{person}{Francisco
  Guzm{\'a}n}, \bibinfo{person}{Edouard Grave}, \bibinfo{person}{Myle Ott},
  \bibinfo{person}{Luke Zettlemoyer}, {and} \bibinfo{person}{Veselin
  Stoyanov}.} \bibinfo{year}{2020}\natexlab{a}.
\newblock \showarticletitle{Unsupervised Cross-lingual Representation Learning
  at Scale}. In \bibinfo{booktitle}{\emph{Proceedings of the 58th Annual
  Meeting of the Association for Computational Linguistics}}.
  \bibinfo{publisher}{Association for Computational Linguistics},
  \bibinfo{address}{Online}, \bibinfo{pages}{8440--8451}.
\newblock
\urldef\tempurl%
\url{https://aclanthology.org/2020.acl-main.747}
\showURL{%
\tempurl}


\bibitem[\protect\citeauthoryear{Conneau, Khandelwal, Goyal, Chaudhary, Wenzek,
  Guzm{\'a}n, Grave, Ott, Zettlemoyer, and Stoyanov}{Conneau
  et~al\mbox{.}}{2020b}]%
        {xlmr}
\bibfield{author}{\bibinfo{person}{Alexis Conneau}, \bibinfo{person}{Kartikay
  Khandelwal}, \bibinfo{person}{Naman Goyal}, \bibinfo{person}{Vishrav
  Chaudhary}, \bibinfo{person}{Guillaume Wenzek}, \bibinfo{person}{Francisco
  Guzm{\'a}n}, \bibinfo{person}{Edouard Grave}, \bibinfo{person}{Myle Ott},
  \bibinfo{person}{Luke Zettlemoyer}, {and} \bibinfo{person}{Veselin
  Stoyanov}.} \bibinfo{year}{2020}\natexlab{b}.
\newblock \showarticletitle{Unsupervised Cross-lingual Representation Learning
  at Scale}. In \bibinfo{booktitle}{\emph{Proceedings of the 58th Annual
  Meeting of the Association for Computational Linguistics}}.
  \bibinfo{publisher}{Association for Computational Linguistics},
  \bibinfo{address}{Online}, \bibinfo{pages}{8440--8451}.
\newblock


\bibitem[\protect\citeauthoryear{Dai and Callan}{Dai and Callan}{2019}]%
        {maxp}
\bibfield{author}{\bibinfo{person}{Zhuyun Dai} {and} \bibinfo{person}{Jamie
  Callan}.} \bibinfo{year}{2019}\natexlab{}.
\newblock \showarticletitle{Deeper text understanding for {IR} with contextual
  neural language modeling}. In \bibinfo{booktitle}{\emph{Proceedings of the
  42nd International ACM SIGIR Conference on Research and Development in
  Information Retrieval}}. \bibinfo{pages}{985--988}.
\newblock


\bibitem[\protect\citeauthoryear{Darwish and Oard}{Darwish and Oard}{2003}]%
        {darwish2003probabilistic}
\bibfield{author}{\bibinfo{person}{Kareem Darwish} {and}
  \bibinfo{person}{Douglas~W Oard}.} \bibinfo{year}{2003}\natexlab{}.
\newblock \showarticletitle{Probabilistic structured query methods}. In
  \bibinfo{booktitle}{\emph{Proceedings of the 26th Annual International ACM
  SIGIR Conference on Research and Development in Information Retrieval}}.
  \bibinfo{pages}{338--344}.
\newblock


\bibitem[\protect\citeauthoryear{Devlin, Chang, Lee, and Toutanova}{Devlin
  et~al\mbox{.}}{2019}]%
        {bert}
\bibfield{author}{\bibinfo{person}{Jacob Devlin}, \bibinfo{person}{Ming-Wei
  Chang}, \bibinfo{person}{Kenton Lee}, {and} \bibinfo{person}{Kristina
  Toutanova}.} \bibinfo{year}{2019}\natexlab{}.
\newblock \showarticletitle{{BERT}: Pre-training of Deep Bidirectional
  Transformers for Language Understanding}. In
  \bibinfo{booktitle}{\emph{Proceedings of the 2019 Conference of the North
  {A}merican Chapter of the Association for Computational Linguistics: Human
  Language Technologies, Volume 1 (Long and Short Papers)}}.
  \bibinfo{publisher}{Association for Computational Linguistics},
  \bibinfo{address}{Minneapolis, Minnesota}, \bibinfo{pages}{4171--4186}.
\newblock


\bibitem[\protect\citeauthoryear{Domhan, Denkowski, Vilar, Niu, Hieber, and
  Heafield}{Domhan et~al\mbox{.}}{2020}]%
        {sockeye2amta}
\bibfield{author}{\bibinfo{person}{Tobias Domhan}, \bibinfo{person}{Michael
  Denkowski}, \bibinfo{person}{David Vilar}, \bibinfo{person}{Xing Niu},
  \bibinfo{person}{Felix Hieber}, {and} \bibinfo{person}{Kenneth Heafield}.}
  \bibinfo{year}{2020}\natexlab{}.
\newblock \showarticletitle{The {Sockeye} 2 Neural Machine Translation Toolkit
  at {AMTA} 2020}. In \bibinfo{booktitle}{\emph{Proceedings of the 14th
  Conference of the Association for Machine Translation in the Americas (Volume
  1: Research Track)}}. \bibinfo{publisher}{Association for Machine Translation
  in the Americas}, \bibinfo{address}{Virtual}, \bibinfo{pages}{110--115}.
\newblock


\bibitem[\protect\citeauthoryear{Formal, Piwowarski, and Clinchant}{Formal
  et~al\mbox{.}}{2021}]%
        {formal2021splade}
\bibfield{author}{\bibinfo{person}{Thibault Formal}, \bibinfo{person}{Benjamin
  Piwowarski}, {and} \bibinfo{person}{St{\'e}phane Clinchant}.}
  \bibinfo{year}{2021}\natexlab{}.
\newblock \showarticletitle{SPLADE: Sparse lexical and expansion model for
  first stage ranking}. In \bibinfo{booktitle}{\emph{Proceedings of the 44th
  International ACM SIGIR Conference on Research and Development in Information
  Retrieval}}. \bibinfo{pages}{2288--2292}.
\newblock


\bibitem[\protect\citeauthoryear{Gao and Callan}{Gao and Callan}{2022}]%
        {cocondenser}
\bibfield{author}{\bibinfo{person}{Luyu Gao} {and} \bibinfo{person}{Jamie
  Callan}.} \bibinfo{year}{2022}\natexlab{}.
\newblock \showarticletitle{Unsupervised Corpus Aware Language Model
  Pre-training for Dense Passage Retrieval}. In
  \bibinfo{booktitle}{\emph{Proceedings of the 60th Annual Meeting of the
  Association for Computational Linguistics (Volume 1: Long Papers)}}.
  \bibinfo{publisher}{Association for Computational Linguistics},
  \bibinfo{address}{Dublin, Ireland}, \bibinfo{pages}{2843--2853}.
\newblock


\bibitem[\protect\citeauthoryear{Granell}{Granell}{2014}]%
        {granell2014multilingual}
\bibfield{author}{\bibinfo{person}{Ximo Granell}.}
  \bibinfo{year}{2014}\natexlab{}.
\newblock \bibinfo{booktitle}{\emph{Multilingual information management:
  Information, technology and translators}}.
\newblock \bibinfo{publisher}{Chandos Publishing}.
\newblock


\bibitem[\protect\citeauthoryear{Hieber, Domhan, Denkowski, and Vilar}{Hieber
  et~al\mbox{.}}{2020}]%
        {sockeye2whitepaper}
\bibfield{author}{\bibinfo{person}{Felix Hieber}, \bibinfo{person}{Tobias
  Domhan}, \bibinfo{person}{Michael Denkowski}, {and} \bibinfo{person}{David
  Vilar}.} \bibinfo{year}{2020}\natexlab{}.
\newblock \showarticletitle{SOCKEYE 2: A Toolkit for Neural Machine
  Translation}. In \bibinfo{booktitle}{\emph{EAMT 2020}}.
\newblock
\urldef\tempurl%
\url{https://www.amazon.science/publications/sockeye-2-a-toolkit-for-neural-machine-translation}
\showURL{%
\tempurl}


\bibitem[\protect\citeauthoryear{Hofst{\"a}tter, Lin, Yang, Lin, and
  Hanbury}{Hofst{\"a}tter et~al\mbox{.}}{2021}]%
        {hofstatter2021efficiently}
\bibfield{author}{\bibinfo{person}{Sebastian Hofst{\"a}tter},
  \bibinfo{person}{Sheng-Chieh Lin}, \bibinfo{person}{Jheng-Hong Yang},
  \bibinfo{person}{Jimmy Lin}, {and} \bibinfo{person}{Allan Hanbury}.}
  \bibinfo{year}{2021}\natexlab{}.
\newblock \showarticletitle{Efficiently teaching an effective dense retriever
  with balanced topic aware sampling}. In \bibinfo{booktitle}{\emph{Proceedings
  of the 44th International ACM SIGIR Conference on Research and Development in
  Information Retrieval}}. \bibinfo{pages}{113--122}.
\newblock


\bibitem[\protect\citeauthoryear{Huang, Zeng, Zamani, and Allan}{Huang
  et~al\mbox{.}}{2023}]%
        {huang2023soft}
\bibfield{author}{\bibinfo{person}{Zhiqi Huang}, \bibinfo{person}{Hansi Zeng},
  \bibinfo{person}{Hamed Zamani}, {and} \bibinfo{person}{James Allan}.}
  \bibinfo{year}{2023}\natexlab{}.
\newblock \showarticletitle{Soft Prompt Decoding for Multilingual Dense
  Retrieval}.
\newblock \bibinfo{journal}{\emph{arXiv preprint arXiv:2305.09025}}
  (\bibinfo{year}{2023}).
\newblock


\bibitem[\protect\citeauthoryear{Hull and Grefenstette}{Hull and
  Grefenstette}{1996}]%
        {hull1996querying}
\bibfield{author}{\bibinfo{person}{David~A Hull} {and} \bibinfo{person}{Gregory
  Grefenstette}.} \bibinfo{year}{1996}\natexlab{}.
\newblock \showarticletitle{Querying across languages: A dictionary-based
  approach to multilingual information retrieval}. In
  \bibinfo{booktitle}{\emph{Proceedings of the 19th Annual International ACM
  SIGIR Conference on Research and Development in Information Retrieval}}.
  \bibinfo{pages}{49--57}.
\newblock


\bibitem[\protect\citeauthoryear{Jegou, Douze, and Schmid}{Jegou
  et~al\mbox{.}}{2010}]%
        {jegou2010product}
\bibfield{author}{\bibinfo{person}{Herve Jegou}, \bibinfo{person}{Matthijs
  Douze}, {and} \bibinfo{person}{Cordelia Schmid}.}
  \bibinfo{year}{2010}\natexlab{}.
\newblock \showarticletitle{Product quantization for nearest neighbor search}.
\newblock \bibinfo{journal}{\emph{IEEE transactions on pattern analysis and
  machine intelligence}} \bibinfo{volume}{33}, \bibinfo{number}{1}
  (\bibinfo{year}{2010}), \bibinfo{pages}{117--128}.
\newblock


\bibitem[\protect\citeauthoryear{Jeronymo, Lotufo, and Nogueira}{Jeronymo
  et~al\mbox{.}}{2023}]%
        {unicamp-at-neuclir}
\bibfield{author}{\bibinfo{person}{Vitor Jeronymo}, \bibinfo{person}{Roberto
  Lotufo}, {and} \bibinfo{person}{Rodrigo Nogueira}.}
  \bibinfo{year}{2023}\natexlab{}.
\newblock \showarticletitle{{NeuralMind-UNICAMP at 2022 TREC NeuCLIR}: Large
  Boring Rerankers for Cross-lingual Retrieval}.
\newblock \bibinfo{journal}{\emph{arXiv preprint arXiv:2303.16145}}
  (\bibinfo{year}{2023}).
\newblock


\bibitem[\protect\citeauthoryear{Johnson, Douze, and J{\'e}gou}{Johnson
  et~al\mbox{.}}{2019}]%
        {faiss}
\bibfield{author}{\bibinfo{person}{Jeff Johnson}, \bibinfo{person}{Matthijs
  Douze}, {and} \bibinfo{person}{Herv{\'e} J{\'e}gou}.}
  \bibinfo{year}{2019}\natexlab{}.
\newblock \showarticletitle{Billion-scale similarity search with {GPUs}}.
\newblock \bibinfo{journal}{\emph{IEEE Transactions on Big Data}}
  \bibinfo{volume}{7}, \bibinfo{number}{3} (\bibinfo{year}{2019}),
  \bibinfo{pages}{535--547}.
\newblock


\bibitem[\protect\citeauthoryear{Karpukhin, O{\u{g}}uz, Min, Lewis, Wu, Edunov,
  Chen, and Yih}{Karpukhin et~al\mbox{.}}{2020}]%
        {dpr}
\bibfield{author}{\bibinfo{person}{Vladimir Karpukhin}, \bibinfo{person}{Barlas
  O{\u{g}}uz}, \bibinfo{person}{Sewon Min}, \bibinfo{person}{Patrick Lewis},
  \bibinfo{person}{Ledell Wu}, \bibinfo{person}{Sergey Edunov},
  \bibinfo{person}{Danqi Chen}, {and} \bibinfo{person}{Wen-tau Yih}.}
  \bibinfo{year}{2020}\natexlab{}.
\newblock \showarticletitle{Dense passage retrieval for open-domain question
  answering}.
\newblock \bibinfo{journal}{\emph{arXiv preprint arXiv:2004.04906}}
  (\bibinfo{year}{2020}).
\newblock


\bibitem[\protect\citeauthoryear{Khattab and Zaharia}{Khattab and
  Zaharia}{2020}]%
        {colbert}
\bibfield{author}{\bibinfo{person}{Omar Khattab} {and} \bibinfo{person}{Matei
  Zaharia}.} \bibinfo{year}{2020}\natexlab{}.
\newblock \showarticletitle{Colbert: Efficient and effective passage search via
  contextualized late interaction over bert}. In
  \bibinfo{booktitle}{\emph{Proceedings of the 43rd International ACM SIGIR
  conference on research and development in Information Retrieval}}.
  \bibinfo{pages}{39--48}.
\newblock


\bibitem[\protect\citeauthoryear{Kraaij, Nie, and Simard}{Kraaij
  et~al\mbox{.}}{2003}]%
        {kraaij2003embedding}
\bibfield{author}{\bibinfo{person}{Wessel Kraaij}, \bibinfo{person}{Jian-Yun
  Nie}, {and} \bibinfo{person}{Michel Simard}.}
  \bibinfo{year}{2003}\natexlab{}.
\newblock \showarticletitle{Embedding web-based statistical translation models
  in cross-language information retrieval}.
\newblock \bibinfo{journal}{\emph{Computational Linguistics}}
  \bibinfo{volume}{29}, \bibinfo{number}{3} (\bibinfo{year}{2003}),
  \bibinfo{pages}{381--419}.
\newblock


\bibitem[\protect\citeauthoryear{Lakens}{Lakens}{2017}]%
        {lakens2017equivalence}
\bibfield{author}{\bibinfo{person}{Dani{\"e}l Lakens}.}
  \bibinfo{year}{2017}\natexlab{}.
\newblock \showarticletitle{Equivalence tests: A practical primer for t tests,
  correlations, and meta-analyses}.
\newblock \bibinfo{journal}{\emph{Social psychological and personality
  science}} \bibinfo{volume}{8}, \bibinfo{number}{4} (\bibinfo{year}{2017}),
  \bibinfo{pages}{355--362}.
\newblock


\bibitem[\protect\citeauthoryear{Lawrie, MacAvaney, Mayfield, McNamee, Oard,
  Soldanini, and Yang}{Lawrie et~al\mbox{.}}{2022a}]%
        {neuclir2022}
\bibfield{author}{\bibinfo{person}{Dawn Lawrie}, \bibinfo{person}{Sean
  MacAvaney}, \bibinfo{person}{James Mayfield}, \bibinfo{person}{Paul McNamee},
  \bibinfo{person}{Douglas~W. Oard}, \bibinfo{person}{Luca Soldanini}, {and}
  \bibinfo{person}{Eugene Yang}.} \bibinfo{year}{2022}\natexlab{a}.
\newblock \showarticletitle{Overview of the TREC 2022 {NeuCLIR} Track}. In
  \bibinfo{booktitle}{\emph{The Thirty-first Text REtrieval Conference (TREC
  2022) Proceedings}}.
\newblock


\bibitem[\protect\citeauthoryear{Lawrie, MacAvaney, Mayfield, McNamee, Oard,
  Soldanini, and Yang}{Lawrie et~al\mbox{.}}{2023a}]%
        {neuclir2023}
\bibfield{author}{\bibinfo{person}{Dawn Lawrie}, \bibinfo{person}{Sean
  MacAvaney}, \bibinfo{person}{James Mayfield}, \bibinfo{person}{Paul McNamee},
  \bibinfo{person}{Douglas~W. Oard}, \bibinfo{person}{Luca Soldanini}, {and}
  \bibinfo{person}{Eugene Yang}.} \bibinfo{year}{2023}\natexlab{a}.
\newblock \showarticletitle{Overview of the TREC 2023 {NeuCLIR} Track}. In
  \bibinfo{booktitle}{\emph{The Thirty-second Text REtrieval Conference (TREC
  2023) Proceedings}}.
\newblock


\bibitem[\protect\citeauthoryear{Lawrie, Mayfield, Oard, and Yang}{Lawrie
  et~al\mbox{.}}{2022b}]%
        {hc4}
\bibfield{author}{\bibinfo{person}{Dawn Lawrie}, \bibinfo{person}{James
  Mayfield}, \bibinfo{person}{Douglas~W. Oard}, {and} \bibinfo{person}{Eugene
  Yang}.} \bibinfo{year}{2022}\natexlab{b}.
\newblock \showarticletitle{{HC}4: A New Suite of Test Collections for Ad Hoc
  {CLIR}}. In \bibinfo{booktitle}{\emph{Proceedings of the 44th European
  Conference on Information Retrieval}}.
\newblock


\bibitem[\protect\citeauthoryear{Lawrie, Yang, Oard, and Mayfield}{Lawrie
  et~al\mbox{.}}{2023b}]%
        {mtt}
\bibfield{author}{\bibinfo{person}{Dawn Lawrie}, \bibinfo{person}{Eugene Yang},
  \bibinfo{person}{Douglas~W Oard}, {and} \bibinfo{person}{James Mayfield}.}
  \bibinfo{year}{2023}\natexlab{b}.
\newblock \showarticletitle{Neural Approaches to Multilingual Information
  Retrieval}. In \bibinfo{booktitle}{\emph{European Conference on Information
  Retrieval}}. Springer, \bibinfo{pages}{521--536}.
\newblock


\bibitem[\protect\citeauthoryear{Li, Lin, Ma, and Lin}{Li
  et~al\mbox{.}}{2023a}]%
        {li2023slim}
\bibfield{author}{\bibinfo{person}{Minghan Li}, \bibinfo{person}{Sheng-Chieh
  Lin}, \bibinfo{person}{Xueguang Ma}, {and} \bibinfo{person}{Jimmy Lin}.}
  \bibinfo{year}{2023}\natexlab{a}.
\newblock \showarticletitle{SLIM: Sparsified Late Interaction for Multi-Vector
  Retrieval with Inverted Indexes}. In \bibinfo{booktitle}{\emph{Proceedings of
  the 46th International ACM SIGIR Conference on Research and Development in
  Information Retrieval}} (, Taipei, Taiwan,) \emph{(\bibinfo{series}{SIGIR
  '23})}. \bibinfo{publisher}{Association for Computing Machinery},
  \bibinfo{address}{New York, NY, USA}, \bibinfo{pages}{1954–1959}.
\newblock
\showISBNx{9781450394086}
\urldef\tempurl%
\url{https://doi.org/10.1145/3539618.3591977}
\showDOI{\tempurl}


\bibitem[\protect\citeauthoryear{Li, Lin, Oguz, Ghoshal, Lin, Mehdad, Yih, and
  Chen}{Li et~al\mbox{.}}{2023b}]%
        {li2023citadel}
\bibfield{author}{\bibinfo{person}{Minghan Li}, \bibinfo{person}{Sheng-Chieh
  Lin}, \bibinfo{person}{Barlas Oguz}, \bibinfo{person}{Asish Ghoshal},
  \bibinfo{person}{Jimmy Lin}, \bibinfo{person}{Yashar Mehdad},
  \bibinfo{person}{Wen-tau Yih}, {and} \bibinfo{person}{Xilun Chen}.}
  \bibinfo{year}{2023}\natexlab{b}.
\newblock \showarticletitle{{CITADEL}: Conditional Token Interaction via
  Dynamic Lexical Routing for Efficient and Effective Multi-Vector Retrieval}.
  In \bibinfo{booktitle}{\emph{Proceedings of the 61st Annual Meeting of the
  Association for Computational Linguistics (Volume 1: Long Papers)}},
  \bibfield{editor}{\bibinfo{person}{Anna Rogers}, \bibinfo{person}{Jordan
  Boyd-Graber}, {and} \bibinfo{person}{Naoaki Okazaki}} (Eds.).
  \bibinfo{publisher}{Association for Computational Linguistics},
  \bibinfo{address}{Toronto, Canada}, \bibinfo{pages}{11891--11907}.
\newblock
\urldef\tempurl%
\url{https://doi.org/10.18653/v1/2023.acl-long.663}
\showDOI{\tempurl}


\bibitem[\protect\citeauthoryear{Li, Franz, Sultan, Iyer, Lee, and Sil}{Li
  et~al\mbox{.}}{2022}]%
        {li2022learning}
\bibfield{author}{\bibinfo{person}{Yulong Li}, \bibinfo{person}{Martin Franz},
  \bibinfo{person}{Md~Arafat Sultan}, \bibinfo{person}{Bhavani Iyer},
  \bibinfo{person}{Young-Suk Lee}, {and} \bibinfo{person}{Avirup Sil}.}
  \bibinfo{year}{2022}\natexlab{}.
\newblock \showarticletitle{Learning Cross-Lingual IR from an {English}
  Retriever}. In \bibinfo{booktitle}{\emph{Proceedings of the 2022 Conference
  of the North American Chapter of the Association for Computational
  Linguistics: Human Language Technologies}}. \bibinfo{pages}{4428--4436}.
\newblock


\bibitem[\protect\citeauthoryear{Magdy and Jones}{Magdy and Jones}{2011}]%
        {magdy2011should}
\bibfield{author}{\bibinfo{person}{Walid Magdy} {and}
  \bibinfo{person}{Gareth~J.F. Jones}.} \bibinfo{year}{2011}\natexlab{}.
\newblock \showarticletitle{Should {MT} systems be used as black boxes in
  {CLIR}?}. In \bibinfo{booktitle}{\emph{European Conference on Information
  Retrieval}}. Springer, \bibinfo{pages}{683--686}.
\newblock


\bibitem[\protect\citeauthoryear{Malkov and Yashunin}{Malkov and
  Yashunin}{2018}]%
        {hnsw}
\bibfield{author}{\bibinfo{person}{Yu~A Malkov} {and} \bibinfo{person}{Dmitry~A
  Yashunin}.} \bibinfo{year}{2018}\natexlab{}.
\newblock \showarticletitle{Efficient and robust approximate nearest neighbor
  search using hierarchical navigable small world graphs}.
\newblock \bibinfo{journal}{\emph{IEEE transactions on pattern analysis and
  machine intelligence}} \bibinfo{volume}{42}, \bibinfo{number}{4}
  (\bibinfo{year}{2018}), \bibinfo{pages}{824--836}.
\newblock


\bibitem[\protect\citeauthoryear{McNamee and Mayfield}{McNamee and
  Mayfield}{2002}]%
        {mcnamee2002comparing}
\bibfield{author}{\bibinfo{person}{Paul McNamee} {and} \bibinfo{person}{James
  Mayfield}.} \bibinfo{year}{2002}\natexlab{}.
\newblock \showarticletitle{Comparing cross-language query expansion techniques
  by degrading translation resources}. In \bibinfo{booktitle}{\emph{Proceedings
  of the 25th annual international ACM SIGIR conference on Research and
  development in information retrieval}}. \bibinfo{pages}{159--166}.
\newblock


\bibitem[\protect\citeauthoryear{Mitamura, Nyberg, Shima, Kato, Mori, Lin,
  Song, Lin, Sakai, Ji, et~al\mbox{.}}{Mitamura et~al\mbox{.}}{2008}]%
        {ntcir2007overview}
\bibfield{author}{\bibinfo{person}{Teruko Mitamura}, \bibinfo{person}{Eric
  Nyberg}, \bibinfo{person}{Hideki Shima}, \bibinfo{person}{Tsuneaki Kato},
  \bibinfo{person}{Tatsunori Mori}, \bibinfo{person}{Chin-Yew Lin},
  \bibinfo{person}{Ruihua Song}, \bibinfo{person}{Chuan-Jie Lin},
  \bibinfo{person}{Tetsuya Sakai}, \bibinfo{person}{Donghong Ji},
  {et~al\mbox{.}}} \bibinfo{year}{2008}\natexlab{}.
\newblock \showarticletitle{Overview of the {NTCIR-7 ACLIA} Tasks: Advanced
  Cross-Lingual Information Access.}. In \bibinfo{booktitle}{\emph{NTCIR}}.
\newblock


\bibitem[\protect\citeauthoryear{Nair, Yang, Lawrie, Duh, McNamee, Murray,
  Mayfield, and Oard}{Nair et~al\mbox{.}}{2022}]%
        {colbertx}
\bibfield{author}{\bibinfo{person}{Suraj Nair}, \bibinfo{person}{Eugene Yang},
  \bibinfo{person}{Dawn Lawrie}, \bibinfo{person}{Kevin Duh},
  \bibinfo{person}{Paul McNamee}, \bibinfo{person}{Kenton Murray},
  \bibinfo{person}{James Mayfield}, {and} \bibinfo{person}{Douglas~W. Oard}.}
  \bibinfo{year}{2022}\natexlab{}.
\newblock \showarticletitle{Transfer Learning Approaches for Building
  Cross-Language Dense Retrieval Models}. In \bibinfo{booktitle}{\emph{Advances
  in Information Retrieval: 44th European Conference on IR Research, ECIR 2022,
  Stavanger, Norway, April 10–14, 2022, Proceedings, Part I}} (Stavanger,
  Norway). \bibinfo{publisher}{Springer-Verlag}, \bibinfo{address}{Berlin,
  Heidelberg}, \bibinfo{pages}{382–396}.
\newblock
\showISBNx{978-3-030-99735-9}


\bibitem[\protect\citeauthoryear{Nair, Yang, Lawrie, Mayfield, and Oard}{Nair
  et~al\mbox{.}}{2023}]%
        {blade}
\bibfield{author}{\bibinfo{person}{Suraj Nair}, \bibinfo{person}{Eugene Yang},
  \bibinfo{person}{Dawn Lawrie}, \bibinfo{person}{James Mayfield}, {and}
  \bibinfo{person}{Douglas~W. Oard}.} \bibinfo{year}{2023}\natexlab{}.
\newblock \showarticletitle{{BLADE}: Combining Vocabulary Pruning and
  Intermediate Pretraining for Scaleable Neural {CLIR}}. In
  \bibinfo{booktitle}{\emph{Proceedings of the 46th International ACM SIGIR
  Conference on Research and Development in Information Retrieval}} (Taipei,
  Taiwan) \emph{(\bibinfo{series}{SIGIR '23})}. \bibinfo{publisher}{Association
  for Computing Machinery}, \bibinfo{address}{New York, NY, USA},
  \bibinfo{pages}{1219–1229}.
\newblock
\showISBNx{9781450394086}


\bibitem[\protect\citeauthoryear{Nguyen, Rosenberg, Song, Gao, Tiwary,
  Majumder, and Deng}{Nguyen et~al\mbox{.}}{2016}]%
        {msmarco}
\bibfield{author}{\bibinfo{person}{Tri Nguyen}, \bibinfo{person}{Mir
  Rosenberg}, \bibinfo{person}{Xia Song}, \bibinfo{person}{Jianfeng Gao},
  \bibinfo{person}{Saurabh Tiwary}, \bibinfo{person}{Rangan Majumder}, {and}
  \bibinfo{person}{Li Deng}.} \bibinfo{year}{2016}\natexlab{}.
\newblock \showarticletitle{{MS} {MARCO:} {A} Human Generated MAchine Reading
  COmprehension Dataset}.
\newblock \bibinfo{journal}{\emph{arXiv preprint arXiv:1611.09268}}
  (\bibinfo{year}{2016}).
\newblock
\showeprint[arXiv]{1611.09268}
\urldef\tempurl%
\url{http://arxiv.org/abs/1611.09268}
\showURL{%
\tempurl}


\bibitem[\protect\citeauthoryear{Nogueira, Jiang, Pradeep, and Lin}{Nogueira
  et~al\mbox{.}}{2020}]%
        {monot5}
\bibfield{author}{\bibinfo{person}{Rodrigo Nogueira}, \bibinfo{person}{Zhiying
  Jiang}, \bibinfo{person}{Ronak Pradeep}, {and} \bibinfo{person}{Jimmy Lin}.}
  \bibinfo{year}{2020}\natexlab{}.
\newblock \showarticletitle{Document Ranking with a Pretrained
  Sequence-to-Sequence Model}. In \bibinfo{booktitle}{\emph{Findings of the
  Association for Computational Linguistics: EMNLP 2020}}.
  \bibinfo{publisher}{Association for Computational Linguistics},
  \bibinfo{address}{Online}, \bibinfo{pages}{708--718}.
\newblock
\urldef\tempurl%
\url{https://doi.org/10.18653/v1/2020.findings-emnlp.63}
\showDOI{\tempurl}


\bibitem[\protect\citeauthoryear{Peters and Braschler}{Peters and
  Braschler}{2002}]%
        {peters2002importance}
\bibfield{author}{\bibinfo{person}{Carol Peters} {and} \bibinfo{person}{Martin
  Braschler}.} \bibinfo{year}{2002}\natexlab{}.
\newblock \showarticletitle{The Importance of Evaluation for Cross-Language
  System Development: the CLEF Experience.}. In
  \bibinfo{booktitle}{\emph{LREC}}.
\newblock


\bibitem[\protect\citeauthoryear{Peters, Braschler, and Clough}{Peters
  et~al\mbox{.}}{2012}]%
        {peters2012multilingual}
\bibfield{author}{\bibinfo{person}{Carol Peters}, \bibinfo{person}{Martin
  Braschler}, {and} \bibinfo{person}{Paul Clough}.}
  \bibinfo{year}{2012}\natexlab{}.
\newblock \bibinfo{booktitle}{\emph{Multilingual information retrieval: From
  research to practice}}.
\newblock \bibinfo{publisher}{Springer}.
\newblock


\bibitem[\protect\citeauthoryear{Qu, Ding, Liu, Liu, Ren, Zhao, Dong, Wu, and
  Wang}{Qu et~al\mbox{.}}{2021}]%
        {rocketqa}
\bibfield{author}{\bibinfo{person}{Yingqi Qu}, \bibinfo{person}{Yuchen Ding},
  \bibinfo{person}{Jing Liu}, \bibinfo{person}{Kai Liu},
  \bibinfo{person}{Ruiyang Ren}, \bibinfo{person}{Wayne~Xin Zhao},
  \bibinfo{person}{Daxiang Dong}, \bibinfo{person}{Hua Wu}, {and}
  \bibinfo{person}{Haifeng Wang}.} \bibinfo{year}{2021}\natexlab{}.
\newblock \showarticletitle{{R}ocket{QA}: An Optimized Training Approach to
  Dense Passage Retrieval for Open-Domain Question Answering}. In
  \bibinfo{booktitle}{\emph{Proceedings of the 2021 Conference of the North
  American Chapter of the Association for Computational Linguistics: Human
  Language Technologies}}. \bibinfo{publisher}{Association for Computational
  Linguistics}, \bibinfo{address}{Online}, \bibinfo{pages}{5835--5847}.
\newblock


\bibitem[\protect\citeauthoryear{Rahimi, Shakery, and King}{Rahimi
  et~al\mbox{.}}{2015}]%
        {rahimi2015multilingual}
\bibfield{author}{\bibinfo{person}{Razieh Rahimi}, \bibinfo{person}{Azadeh
  Shakery}, {and} \bibinfo{person}{Irwin King}.}
  \bibinfo{year}{2015}\natexlab{}.
\newblock \showarticletitle{Multilingual information retrieval in the language
  modeling framework}.
\newblock \bibinfo{journal}{\emph{Information Retrieval Journal}}
  \bibinfo{volume}{18}, \bibinfo{number}{3} (\bibinfo{year}{2015}),
  \bibinfo{pages}{246--281}.
\newblock


\bibitem[\protect\citeauthoryear{Reimers and Gurevych}{Reimers and
  Gurevych}{2019}]%
        {reimers2019sentence}
\bibfield{author}{\bibinfo{person}{Nils Reimers} {and} \bibinfo{person}{Iryna
  Gurevych}.} \bibinfo{year}{2019}\natexlab{}.
\newblock \showarticletitle{Sentence-BERT: Sentence Embeddings using Siamese
  BERT-Networks}. In \bibinfo{booktitle}{\emph{Proceedings of the 2019
  Conference on Empirical Methods in Natural Language Processing and the 9th
  International Joint Conference on Natural Language Processing
  (EMNLP-IJCNLP)}}. Association for Computational Linguistics.
\newblock


\bibitem[\protect\citeauthoryear{Santhanam, Khattab, Saad-Falcon, Potts, and
  Zaharia}{Santhanam et~al\mbox{.}}{2022}]%
        {colbertv2}
\bibfield{author}{\bibinfo{person}{Keshav Santhanam}, \bibinfo{person}{Omar
  Khattab}, \bibinfo{person}{Jon Saad-Falcon}, \bibinfo{person}{Christopher
  Potts}, {and} \bibinfo{person}{Matei Zaharia}.}
  \bibinfo{year}{2022}\natexlab{}.
\newblock \showarticletitle{{C}ol{BERT}v2: Effective and Efficient Retrieval
  via Lightweight Late Interaction}. In \bibinfo{booktitle}{\emph{Proceedings
  of the 2022 Conference of the North American Chapter of the Association for
  Computational Linguistics: Human Language Technologies}}.
  \bibinfo{publisher}{Association for Computational Linguistics},
  \bibinfo{address}{Seattle, United States}, \bibinfo{pages}{3715--3734}.
\newblock


\bibitem[\protect\citeauthoryear{Schuirmann}{Schuirmann}{1987}]%
        {schuirmann1987comparison}
\bibfield{author}{\bibinfo{person}{Donald~J Schuirmann}.}
  \bibinfo{year}{1987}\natexlab{}.
\newblock \showarticletitle{A comparison of the two one-sided tests procedure
  and the power approach for assessing the equivalence of average
  bioavailability}.
\newblock \bibinfo{journal}{\emph{Journal of pharmacokinetics and
  biopharmaceutics}}  \bibinfo{volume}{15} (\bibinfo{year}{1987}),
  \bibinfo{pages}{657--680}.
\newblock


\bibitem[\protect\citeauthoryear{Shi and Lin}{Shi and Lin}{2019}]%
        {shi2019cross}
\bibfield{author}{\bibinfo{person}{P Shi} {and} \bibinfo{person}{J Lin}.}
  \bibinfo{year}{2019}\natexlab{}.
\newblock \showarticletitle{Cross-lingual relevance transfer for document
  retrieval}.
\newblock \bibinfo{journal}{\emph{arXiv preprint arXiv:1911.02989}}
  (\bibinfo{year}{2019}).
\newblock


\bibitem[\protect\citeauthoryear{Si, Callan, Cetintas, and Yuan}{Si
  et~al\mbox{.}}{2008}]%
        {si2008effective}
\bibfield{author}{\bibinfo{person}{Luo Si}, \bibinfo{person}{Jamie Callan},
  \bibinfo{person}{Suleyman Cetintas}, {and} \bibinfo{person}{Hao Yuan}.}
  \bibinfo{year}{2008}\natexlab{}.
\newblock \showarticletitle{An effective and efficient results merging strategy
  for multilingual information retrieval in federated search environments}.
\newblock \bibinfo{journal}{\emph{Information Retrieval}} \bibinfo{volume}{11},
  \bibinfo{number}{1} (\bibinfo{year}{2008}), \bibinfo{pages}{1--24}.
\newblock


\bibitem[\protect\citeauthoryear{Tsai, Wang, and Chen}{Tsai
  et~al\mbox{.}}{2008}]%
        {tsai2008study}
\bibfield{author}{\bibinfo{person}{Ming-Feng Tsai}, \bibinfo{person}{Yu-Ting
  Wang}, {and} \bibinfo{person}{Hsin-Hsi Chen}.}
  \bibinfo{year}{2008}\natexlab{}.
\newblock \showarticletitle{A study of learning a merge model for multilingual
  information retrieval}. In \bibinfo{booktitle}{\emph{Proceedings of the 31st
  Annual International ACM SIGIR Conference on Research and Development in
  Information Retrieval}}. \bibinfo{pages}{195--202}.
\newblock


\bibitem[\protect\citeauthoryear{Xu and Weischedel}{Xu and Weischedel}{2000}]%
        {xu2000cross}
\bibfield{author}{\bibinfo{person}{Jinxi Xu} {and} \bibinfo{person}{Ralph
  Weischedel}.} \bibinfo{year}{2000}\natexlab{}.
\newblock \showarticletitle{Cross-lingual information retrieval using hidden
  {Markov} models}. In \bibinfo{booktitle}{\emph{2000 Joint SIGDAT Conference
  on Empirical Methods in Natural Language Processing and Very Large Corpora}}.
  \bibinfo{pages}{95--103}.
\newblock


\bibitem[\protect\citeauthoryear{Xue, Constant, Roberts, Kale, Al-Rfou,
  Siddhant, Barua, and Raffel}{Xue et~al\mbox{.}}{2021}]%
        {mt5}
\bibfield{author}{\bibinfo{person}{Linting Xue}, \bibinfo{person}{Noah
  Constant}, \bibinfo{person}{Adam Roberts}, \bibinfo{person}{Mihir Kale},
  \bibinfo{person}{Rami Al-Rfou}, \bibinfo{person}{Aditya Siddhant},
  \bibinfo{person}{Aditya Barua}, {and} \bibinfo{person}{Colin Raffel}.}
  \bibinfo{year}{2021}\natexlab{}.
\newblock \showarticletitle{m{T}5: A Massively Multilingual Pre-trained
  Text-to-Text Transformer}. In \bibinfo{booktitle}{\emph{Proceedings of the
  2021 Conference of the North American Chapter of the Association for
  Computational Linguistics: Human Language Technologies}}.
  \bibinfo{publisher}{Association for Computational Linguistics},
  \bibinfo{address}{Online}, \bibinfo{pages}{483--498}.
\newblock
\urldef\tempurl%
\url{https://doi.org/10.18653/v1/2021.naacl-main.41}
\showDOI{\tempurl}


\bibitem[\protect\citeauthoryear{Yang, Lawrie, Mayfield, Oard, and Miller}{Yang
  et~al\mbox{.}}{2024a}]%
        {tdistill}
\bibfield{author}{\bibinfo{person}{Eugene Yang}, \bibinfo{person}{Dawn Lawrie},
  \bibinfo{person}{James Mayfield}, \bibinfo{person}{Douglas~W Oard}, {and}
  \bibinfo{person}{Scott Miller}.} \bibinfo{year}{2024}\natexlab{a}.
\newblock \showarticletitle{Translate-Distill: Learning Cross-Language Dense
  Retrieval by Translation and Distillation}. In
  \bibinfo{booktitle}{\emph{Advances in Information Retrieval: 46th European
  Conference on IR Research, ECIR 2024}}.
\newblock


\bibitem[\protect\citeauthoryear{Yang, Nair, Chandradevan, Iglesias-Flores, and
  Oard}{Yang et~al\mbox{.}}{2022}]%
        {c3}
\bibfield{author}{\bibinfo{person}{Eugene Yang}, \bibinfo{person}{Suraj Nair},
  \bibinfo{person}{Ramraj Chandradevan}, \bibinfo{person}{Rebecca
  Iglesias-Flores}, {and} \bibinfo{person}{Douglas~W. Oard}.}
  \bibinfo{year}{2022}\natexlab{}.
\newblock \showarticletitle{C3: Continued Pretraining with Contrastive Weak
  Supervision for Cross Language Ad-Hoc Retrieval}. In
  \bibinfo{booktitle}{\emph{Proceedings of the 45th International ACM SIGIR
  Conference on Research and Development in Information Retrieval}} (Madrid,
  Spain) \emph{(\bibinfo{series}{SIGIR '22})}. \bibinfo{publisher}{Association
  for Computing Machinery}, \bibinfo{address}{New York, NY, USA},
  \bibinfo{pages}{2507–2512}.
\newblock
\urldef\tempurl%
\url{https://doi.org/10.1145/3477495.3531886}
\showDOI{\tempurl}


\bibitem[\protect\citeauthoryear{Yang, Nair, Lawrie, Mayfield, Oard, and
  Duh}{Yang et~al\mbox{.}}{2024b}]%
        {yang2024psq}
\bibfield{author}{\bibinfo{person}{Eugene Yang}, \bibinfo{person}{Suraj Nair},
  \bibinfo{person}{Dawn Lawrie}, \bibinfo{person}{James Mayfield},
  \bibinfo{person}{Douglas~W Oard}, {and} \bibinfo{person}{Kevin Duh}.}
  \bibinfo{year}{2024}\natexlab{b}.
\newblock \showarticletitle{Efficiency-Effectiveness Tradeoff of Probabilistic
  Structured Queries for Cross-Language Information Retrieval}.
\newblock \bibinfo{journal}{\emph{arXiv preprint arXiv:2404.18797}}
  (\bibinfo{year}{2024}).
\newblock
\urldef\tempurl%
\url{https://arxiv.org/abs/2404.18797}
\showURL{%
\tempurl}


\bibitem[\protect\citeauthoryear{Zhang, Ma, Shi, and Lin}{Zhang
  et~al\mbox{.}}{2021}]%
        {zhang2021mrtydi}
\bibfield{author}{\bibinfo{person}{Xinyu Zhang}, \bibinfo{person}{Xueguang Ma},
  \bibinfo{person}{Peng Shi}, {and} \bibinfo{person}{Jimmy Lin}.}
  \bibinfo{year}{2021}\natexlab{}.
\newblock \showarticletitle{Mr. {T}y{D}i: A Multi-lingual Benchmark for Dense
  Retrieval}. In \bibinfo{booktitle}{\emph{Proceedings of the 1st Workshop on
  Multilingual Representation Learning}}. \bibinfo{publisher}{Association for
  Computational Linguistics}, \bibinfo{address}{Punta Cana, Dominican
  Republic}, \bibinfo{pages}{127--137}.
\newblock
\urldef\tempurl%
\url{https://aclanthology.org/2021.mrl-1.12}
\showURL{%
\tempurl}


\end{thebibliography}
